\documentstyle[twoside,fleqn,espcrc2,epsfig,amssymb]{article}

\def\lsim{\mathrel{\rlap{\lower4pt\hbox{\hskip1pt$\sim$}}
    \raise1pt\hbox{$<$}}}                
\def\gsim{\mathrel{\rlap{\lower4pt\hbox{\hskip1pt$\sim$}}
    \raise1pt\hbox{$>$}}}                

\def\3{\ss}

\newcommand{\AmS}{{\protect\the\textfont2
  A\kern-.1667em\lower.5ex\hbox{M}\kern-.125emS}}

\title{
       \vspace{-4.00cm}                                     
       {\normalsize DESY 98--153}    \\[-0.2cm]             
       {\normalsize FU-HEP/98--4}    \\[-0.2cm]             
       {\normalsize HUB--EP--98/61}  \\[-0.2cm]             
       {\normalsize TPR-98-30}       \\[-0.2cm]             
       {\normalsize September 1998}  \\                     
       \vspace{2.25cm}                                      
       Mass Spectrum and Decay Constants in the Continuum Limit%
            \thanks{Talk given by D.~Pleiter at Lat98,      
                    Boulder, U.S.A.}}                       

\author{M.~G\"ockeler%
           \address{Institut f\"ur Theoretische Physik, Universit\"at
                    Regensburg, D-93040 Regensburg, Germany},
        R.~Horsley%
           \address{Institut f\"ur Physik, Humboldt-Universit\"at zu Berlin,
                    D-10115 Berlin, Germany},
        V.~Linke%
           \address{Institut f\"ur Theoretische Physik,
                    Freie Universit\"at Berlin, D-14195 Berlin, Germany},
        D.~Pleiter$^{\rm c,}$%
           \address{Deutsches Elektronen-Synchrotron DESY and NIC,
                    D-15735 Zeuthen, Germany},
        P.~E.~L.~Rakow$^{\rm a}$,
        G.~Schierholz$^{\rm d,}$%
           \address{Deutsches Elektronen-Synchrotron DESY,
                    D-22603 Hamburg, Germany},
        A.~Schiller%
           \address{Institut f\"ur Theoretische Physik, Universit\"at
                    Leipzig, D-04109 Leipzig, Germany},
        P.~Stephenson%
           \address{Dipartimento di Fisica,
                    Universit\`a degli Studi di Pisa e INFN,
                    Sezione di Pisa, 56100 Pisa, Italy}
        and
        H.~St\"uben%
           \address{Konrad-Zuse-Zentrum f\"ur Informationstechnik,
                    D-14195 Berlin, Germany}
}
       
\begin{document}

\begin{abstract}
We present first results for light hadron masses, quark masses and
decay constants in the continuum limit using $O(a)$ improved fermions
at three different values of the gauge coupling $\beta$.
\end{abstract}

\maketitle

\setcounter{footnote}{0}


\section{INTRODUCTION}

It has become standard to reduce, or indeed eliminate completely,
the $O(a)$ cut-off effects of Wilson
fermions by adding the local Sheikholeslami-Wohlert counterterm.
%
%
The QCDSF
collaboration has simulated quenched QCD with improved fermions at
three different values of the gauge coupling $\beta=6.0$, $6.2$, and
$6.4$, which covers the range of lattice spacings $a^{-1} \approx
2-3.5$ GeV. The simulations have been done on lattices of size
$16^3 \times 32$ ($\beta=6.0$), $24^3 \times 48$ ($\beta=6.0$, $6.2$),
$32^3 \times 64$ ($\beta=6.2$), and $32^3 \times 48$ ($\beta=6.4$).
The spatial size of the lattices varies between $1.7-2.5$ fm.
We have generated $O(200-1000)$, $O(300)$, $O(100)$ gauge field
configurations
at $\beta=6.0$, $6.2$ and $6.4$, respectively.  They have been evaluated
for 3-8 different values of the hopping parameter $\kappa$ with
$m_\pi/m_\rho$ approximately in the range of $0.4-0.9$.%
\footnote{For more simulation details, see \cite{goeckeler97a}.}
The coefficient of the Sheikholeslami-Wohlert counterterm, $c_{SW}(g^2)$,
was taken from \cite{luescher97a}.
We are currently improving both our statistics and number of $\kappa$
values at $\beta=6.4$.  Our results presented here should
therefore be regarded as preliminary.

\section{CHIRAL EXTRAPOLATION}

\begin{figure}[t]
\vspace*{0.25cm}
\includegraphics[height=5.9cm]{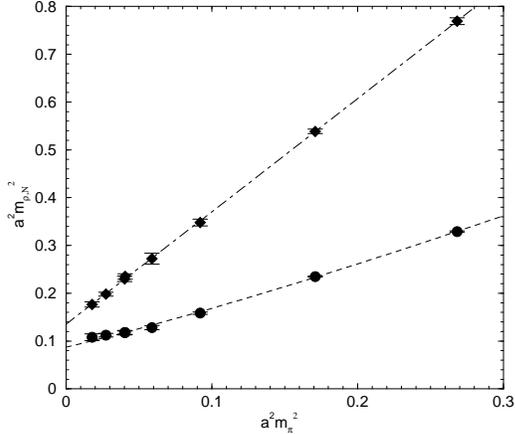}
\vspace*{-1.0cm}
\caption{\footnotesize Chiral extrapolation of $\rho$ ({\large$\bullet$})
         and nucleon masses ($\blacklozenge$) at $\beta=6.2$.}
\vspace*{-0.4cm}
\label{fig:chiral}
\end{figure}
In order to extrapolate the masses and decay constants to the chiral
limit, we use the phenomenological ansatz
\begin{equation}
m_X^2 = b_0 + b_2 m_\pi^2 + b_3 m_\pi^3.
\label{eq:chiral}
\end{equation}
Fits with only three parameters using this ansatz give smaller
$\chi^2/dof$ than the ansatz based on predictions of chiral
perturbation theory, e.g. for $m_\rho$ and $m_N$ \cite{qchiralPT}.
Fig.~\ref{fig:chiral} shows $m_\rho^2$ and $m_N^2$ as a function
of $m_\pi^2$ at $\beta=6.2$.
We have used a two parameter fit (keeping $b_3=0$ fixed) at
$\beta=6.4$, since we currently have results for three $\kappa$ values only.

\begin{figure}[hb]
\vspace*{-0.6cm}
\begin{centering}
\includegraphics[height=5.9cm]{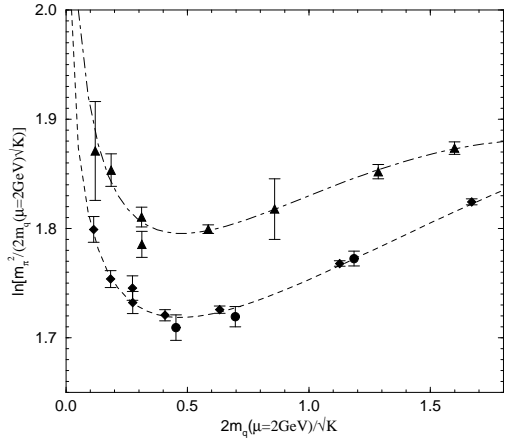}
\end{centering}
\vspace*{-1.0cm}
\caption{\footnotesize Searching for chiral logarithms at
         $\beta=6.0$ ($\blacktriangle$), $\beta=6.2$ ($\blacklozenge$)
         and $\beta=6.4$ ({\large$\bullet$}, not used for fits).}
\label{fig:delta}
\end{figure}
Searching for quenched chiral logarithms predicted by the chiral perturbation
theory we looked at the logarithm of the ratio $m_\pi^2/m_q$
as function of the quark mass $m_q$ \cite{sharpe93a}.
Since the standard method
for determination of $\kappa_c$ depends on the presence of these
singularities, we use the quark masses determined by the Ward identity
method.
Using the ansatz \cite{gupta94a}
\begin{eqnarray}
\label{eq:chirallog}
\ln\frac{(a m_\pi)^2}{a m_q} & = & c_0 - \frac{\delta}{1+\delta} \ln(a m_q) + \\
                             &   & c_1\;a m_q + c_2\;(a m_q)^2 \nonumber
\end{eqnarray}
we find $\delta \approx 0.1-0.2$ for $\beta=6.0$ and $6.2$.
Our data and the fits are plotted in fig.~\ref{fig:delta}.

\section{HADRON MASSES}

\begin{figure}[h]
\vspace*{-0.5cm}
\includegraphics[height=5.9cm]{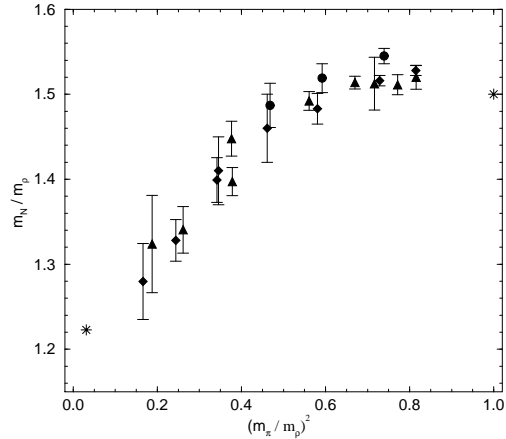}
\vspace*{-0.9cm}
\caption{\footnotesize APE plot for improved fermions at $\beta=6.0$
         ($\blacktriangle$), $\beta=6.2$ ($\blacklozenge$)
         and $\beta=6.4$ ({\large$\bullet$}).
         This is compared with the physical
         mass ratio and the heavy quark limit ($*$). The errors are
         bootstrap errors.}
\vspace*{-0.5cm}
\label{fig:ape}
\end{figure}

In fig.~\ref{fig:ape} we plot our results for the dimensionless mass
ratio $m_N/m_\rho$ versus $(m_\pi/m_\rho)^2$.
Within errors there are no visible cut-off effects. For
$\beta=6.0$ and $6.2$ we find the ratio $m_N/m_\rho$ in the chiral
limit to be in agreement with the experimental value.

%

\begin{figure}[t]
\vspace*{0.38cm}
\includegraphics[height=8.1cm]{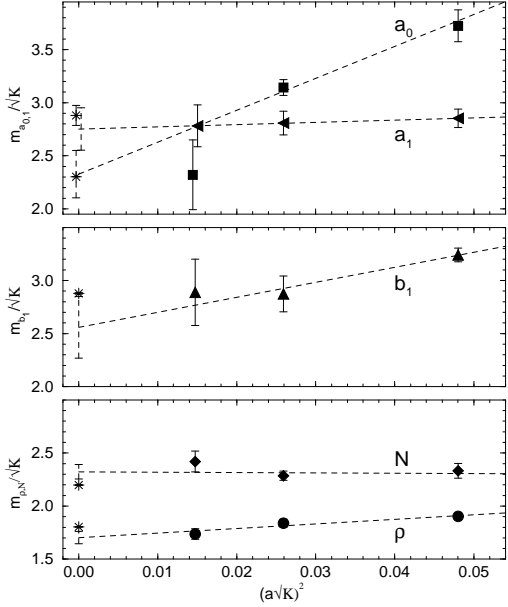}
\vspace*{-1.0cm}
\caption{\footnotesize Continuum extrapolation of various hadron masses
         compared with the experiment ($*$).}
\vspace*{-0.5cm}
\label{fig:spectrum}
\end{figure}

We use a linear fit in $a^2$ in order to extrapolate our results for
the hadron masses to the continuum limit. We use the string tension
$K$, which has cut-off errors of $O(a^2)$, to fix the scale and
$\sqrt{K} = 427 \mbox{MeV}$ to express the experimental values in terms
of $\sqrt{K}$.  As shown in fig.~\ref{fig:spectrum}
our preliminary data agrees with the experimental values.
For the $a_0$ and $b_1$ mesons we find large $O(a^2)$ effects.

\section{QUARK MASSES AND DECAY CONSTANTS}

\begin{figure}[bht]
\vspace*{0.25cm}
\includegraphics[height=5.9cm]{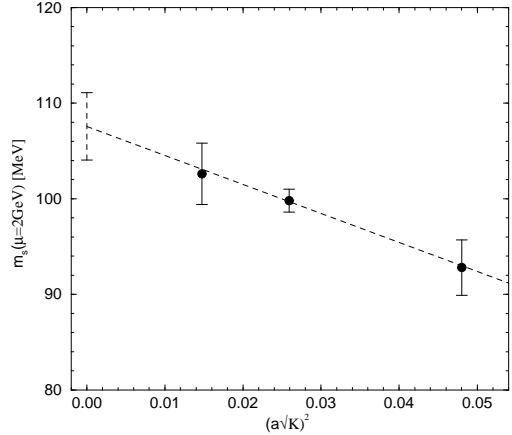}
\vspace*{-1.0cm}
\caption{\footnotesize Continuum extrapolation of the strange quark mass
         $m_s^{\overline{MS}}(2\mbox{GeV})$.}
\label{fig:quark}
\end{figure}
We define the bare quark mass using the Ward identity method.
%
%
While the renormalization constant $Z_A$ and the coefficient
$c_A$ are known non-perturbatively \cite{luescher97a}, we use tadpole
improved values for $Z^{\overline{MS}}_P(a\mu)$, $b_A$ and
$b_P$ \cite{goeckeler97a,capitani97a}. To determine $\hat{m}_s$
we proceed as described in \cite{goeckeler97a} and use the
physical pion and kaon masses as input.

To calculate the decay constants $f_\pi$ and $1/f_\rho$
we use the improved renormalized operators ${\cal A}_\mu =
(1 + b_A a m) Z_A (A_\mu + c_A a \partial_\mu P)$ and
${\cal V}_\mu = (1 + b_V a m) Z_V (V_\mu + i c_V a \partial_\lambda
T_{\mu\lambda})$. $Z_V$, $b_V$ and $c_V$ are known
non-perturbatively \cite{luescher97a}. While $f_\pi$ scales very well,
this is less obvious for $1/f_\rho$. For other matrix elements using
improved axial and vector currents see \cite{roger98a}.
\begin{figure}[th]
\vspace*{-0.2cm}
\includegraphics[height=5.7cm]{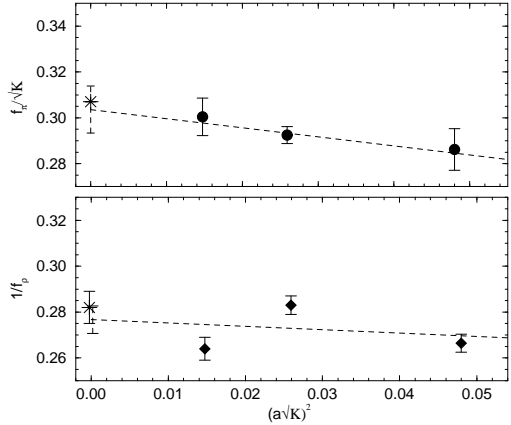}
\vspace*{-1.0cm}
\caption{\footnotesize Decay constants $f_\pi$ and $1/f_\rho$.}
\vspace*{-0.2cm}
\label{fig:decay}
\end{figure}

\section*{ACKNOWLEDGEMENTS}
\label{acknowledgement}

The numerical calculations were performed on the
Ape $QH2$ at DESY-Zeuthen and the Cray $T3D$/$T3E$ at ZIB, Berlin.


\begin{thebibliography}{9}
\bibitem{goeckeler97a}	M. G\"ockeler et al., Phys. Rev. D57 (1998) 5562,
			hep-lat/9707021.
\bibitem{luescher97a}	M. L\"uscher et al.,
			Nucl.Phys. B491 (1997) 323, hep-lat/9609035;
			M. L\"uscher et al.,
			Nucl. Phys. B491 (1997) 344, hep-lat/9611015;
			M. Guagnelli, R. Sommer,
			Nucl. Phys. Proc. Suppl. 63 (1998) 886, hep-lat/9709088;
			S. Sint, P. Weisz,
			Nucl. Phys. B502 (1997) 251, hep-lat/9704001.
\bibitem{qchiralPT}	M. Booth et al.,
			Phys. Rev. D55 (1997) 3092, hep-ph/9610532;
			J. Labrenz, S. Sharpe,
			Nucl. Phys. Proc. Suppl. 34 (1994) 335, hep-lat/9312067.
\bibitem{sharpe93a}	S. Sharpe,
			Nucl. Phys. Proc. Suppl. 30 (1993) 213, hep-lat/9211005.
\bibitem{gupta94a}	R. Gupta,
			Nucl. Phys. Proc. Suppl. 42 (1995) 85, hep-lat/9412078.
\bibitem{capitani97a}	S. Capitani et al.,
			Nucl. Phys. Proc. Suppl. 63 (1998) 874, hep-lat/9709049.
\bibitem{roger98a}	R. Horsley, this conference.
\end{thebibliography}
\end{document}